# Highly tunable Gilbert damping in two-dimensional van der Waals ferromagnet Fe$_3$GaTe$_2$: From bilayer to the twisted bilayer

Jie Wang[1], Shi-Bo Zhao[1], Jia-wan Li[1], Lin Zhuang[1] and Yusheng Hou[1,*]

[1] Guangdong Provincial Key Laboratory of Magnetoelectric Physics and Devices, Center for Neutron Science and Technology, School of Physics, Sun Yat-Sen University, Guangzhou, 510275, China

**ABSTRACT**

Van der Waals ferromagnet Fe$_3$GaTe$_2$ possesses both a high Curie temperature and robust perpendicular magnetic anisotropy, holding promise for practical spintronic applications. In particular, understanding and engineering its Gilbert damping which determines magnetization dynamics are crucial for its applications. Here, we investigate the Gilbert damping of bilayer and the twisted bilayer Fe$_3$GaTe$_2$ through first-principles calculations. For the bilayer Fe$_3$GaTe$_2$, we obtain a quite low Gilbert damping when its magnetization is along the *z* axis at room temperature. In addition, the bilayer Fe$_3$GaTe$_2$ exhibits a large orientational anisotropy of Gilbert damping when its magnetization is rotated from the magnetic easy axis to the hard one. Such anisotropy is attributed to the distinct band structures caused by the anisotropic spin-orbit coupling. Surprisingly, we find that twisting the bilayer Fe$_3$GaTe$_2$ can effectively reduce the Gilbert damping for the perpendicular magnetization, and enhance the orientational anisotropy of Gilbert damping up to 635% when rotating the magnetization from the magnetic easy axis to the hard one. These findings open up an entirely new avenue for the manipulation of Gilbert damping and its anisotropy in two-dimensional van der Waals ferromagnets.

[*]Corresponding authors: houysh@mail.sysu.edu.cn

## I. INTRODUCTION

Since the groundbreaking exfoliation of single-layer graphene [1], the realm of two-dimensional (2D) materials has flourished over the past two decades and garnered intense interest thanks to their unique structures, exceptional properties, and broad application prospects [2-6]. In particular, 2D magnetic materials have been remaining a focal point of research. According to the well-known Mermin-Wagner theorem [7], long-range magnetic orders in 2D magnetic materials are typically precluded by thermal fluctuations. Yet, a perpendicular magnetocrystalline anisotropy, stemming from spin-orbit coupling (SOC), can stabilize magnetic orders by creating a spin wave excitation gap [8]. The pivotal observations of long-range magnetic orders in 2D van der Waals (vdW) crystals like $CrI_3$ monolayer and $Cr_2Ge_2Te_6$ bilayer [8,9] in 2017 significantly catalyzed deeper explorations in this domain. Recently, novel 2D magnetic vdW materials [10-14], such as metallic $Fe_3GeTe_2$ which has gate-tunable room-temperature ferromagnetism in its thin flakes and insulating $CrBr_3$ which exhibits the stacking-dependent interlayer magnetism in its bilayer have emerged, offering new platforms for studying 2D magnetism and designing applications in spintronic devices. Among these 2D vdW magnets, metallic ferromagnetic (FM) $Fe_3GaTe_2$ (FGT) has attracted widespread attentions, due to its high $T_c$ above room-temperature and the robust large room-temperature perpendicular magnetic anisotropy [15] which are highly desirable for practical applications in spintronics.

For the spintronic applications of ferromagnets such as abovementioned FGT, their magnetization dynamics plays a crucial role. Physically, the magnetization evolution of ferromagnets over time is described by the Landau-Lifshitz-Gilbert equation [16]:

$$\frac{d\mathbf{M}}{dt} = -\gamma \mathbf{M} \times \mathbf{H}_{\text{eff}} + \frac{\alpha}{M_S} \mathbf{M} \times \frac{d\mathbf{M}}{dt} \qquad (1).$$

The first term denotes the precession of magnetization due to the effective magnetic field $\mathbf{H}_{\text{eff}}$, and $\gamma = g\mu_0\mu_B/\hbar$ is the gyromagnetic ratio. The second term describes the energy dissipation of this precession, and the dimensionless parameter $\alpha$ is the Gilbert damping parameter [17-19]. Usually, a stronger Gilbert damping enables magnetization to stabilize more rapidly in an external field, while a weaker Gilbert damping means lower energy consumption and longer spin wave propagation [20]. Many functionalities of spintronic devices are decided by the Gilbert damping, such as the signal-to-noise ratio and response speed in hard drives [21,22], the current-induced magnetization switching time in the magnetic random access memories and the velocity of domain wall motion in the race-track memories [23-26]. Hence, determining Gilbert damping and understanding its mechanism are essential to comprehending the magnetization dynamics in ferromagnets and developing promising spintronic devices.

Although Gilbert damping is commonly regarded as a scalar, it has been already demonstrated that it can be an anisotropic tensor in ferromagnets [27-30]. When the magnetization direction varies, Gilbert damping may vary as well. Even along the same direction, Gilbert damping differs at distinct positions in the precession trajectory. These two scenarios are respectively known as orientational and rotational anisotropy of Gilbert damping [28]. That is to say, apart from the conventional temperature effects, Gilbert damping also depends on the orientation of magnetizations. Recently, several experimental studies [31-34] have detected anisotropic Gilbert damping, such as in Fe/α-GeTe bilayer structures and at the interface of Fe/GaAs (001). In addition, non-monotonic thickness-dependent Gilbert damping anisotropy is theoretically found in Bi/Fe heterostructures [35]. These findings demonstrate the complex anisotropic nature of Gilbert damping. Due to the perpendicular magnetic anisotropy, vdW ferromagnet FGT exhibits an energy difference and inherent different symmetries between distinct magnetization orientations, offering a good playground to explore the anisotropy of Gilbert damping.

Given the presence of an additional twist degree of freedom in 2D layered materials, the artificial moiré superlattices formed by vertically stacking their two monolayers and rotating one of the monolayers with a finite twist angle could lead to new fascinating phenomena [36]. In such moiré superlattice, electrons are subject to an extra moiré periodic potential which significantly alters electronic band structures. Meanwhile, the interlayer electronic coupling is changed by twist, which affects the electronic structures and may trigger novel physics [37-42]. For instance, twisted bilayer graphene with the magic angle of 1.1° displays superconductivity or insulating behavior [43,44]. Besides, twist could also cause magnetism and Chern insulators [45-49]. Interestingly, a twisted-angle-sensitive tunnel-magnetoresistance like behavior is experimentally observed in twisted homojunctions of $Fe_3GeTe_2$ with large angles [50]. To date, no studies have been reported on how twist influences Gilbert damping, either experimentally or theoretically. Because vdW layered FGT has inherent interfaces and shares the same symmetry with $Fe_3GeTe_2$, it can be subjected to twist operations and its magnetic properties are likely to change accordingly. Therefore, FGT bilayer is an appealing platform for exploring the impact of twist on Gilbert damping.

In this work, we systematically study the Gilbert damping of bilayer and twisted bilayer FGT employing the first-principles calculations. Our calculations indicate that bilayer FGT has a fairly low Gilbert damping of $2.5\times10^{-3}$ when its magnetization is along its magnetic easy axis (i.e., the *z* axis) at room temperature. Besides, its Gilbert damping exhibits a large orientational anisotropy when its magnetization is rotated from *z* to *x* axis and remains nearly isotropic in the *x-o-y* plane. When bilayer FGT is twisted with an angle of 43.9°, the Gilbert damping with the magnetization along the *z* axis is obviously reduced whereas the Gilbert damping with in-plane magnetizations is enhanced. When the magnetization evolves in the *z-o-x* plane, twist markedly enlarges the anisotropy of Gilbert damping, but reduces the anisotropy of Gilbert damping in the *x-o-y* plane. By examining band structures and the corresponding *k*-dependent contributions to Gilbert damping, we demonstrate that the magnetization orientation dependent SOC leads to anisotropic Gilbert damping in bilayer and twisted bilayer FGT. Our work provides deep insights into the Gilbert damping of 2D vdW ferromagnet FGT and establish twist as an intriguing method for the regulation of Gilbert damping in ultrathin spintronic devices.

## II. COMPUTATIONAL METHODS

Our density functional theory (DFT) calculations are performed with the Vienna *Ab initio* Simulation Package (VASP) [51-53]. To obtain results that are in agreement with experimental data [15], we test the Perdew-Burke-Ernzerhof (PBE) functional combining with DFT-D2 or DFT-D3 and the local density approximation (LDA) functional for structural and magnetic properties of bulk FGT [54-58], as shown in Table S1 in the Supplemental Material (SM) [59]. These results suggest that the LDA functional can more precisely reproduce the structural and magnetic properties of bulk FGT. Given its effectiveness in bulk FGT, the LDA functional is utilized to study the properties of bilayer and twisted bilayer FGT hereafter. To balance computational costs, we select the 43.9° twisted configuration, whose unit cell has 150 atoms totally. Other structures constructed during the twisting process are also provided in the Table S2 of the SM [59].

Based on the scattering theory of Gilbert damping, we adopt an extended torque method to calculate Gilbert damping parameter [60-65], as shown in the following equation:

$$\alpha_{\mu\nu} = -\frac{\pi\hbar\gamma}{M_S}\sum_{ij}\langle\psi_i\left|\frac{\partial H}{\partial u_\mu}\right|\psi_j\rangle\langle\psi_j\left|\frac{\partial H}{\partial u_\nu}\right|\psi_i\rangle \times \delta(E_F - E_i)\delta(E_F - E_j) \quad (2).$$

Some details of Eq. (2) are given in the Part 2 in SM. In our practical numerical calculations, the delta function $\delta(E_F - E)$ in Eq. (2) is replaced by the Lorentzian function $L(E) = 0.5\Gamma/[\pi(E - E_0)^2 + \pi(0.5\Gamma)^2]$ with the scattering rate $\Gamma$ characterizing the temperature effect [62]. To investigate the anisotropy of Gilbert damping in bilayer and twisted bilayer FGT, we define the *a* axis (i.e. *x* axis) as the reference and compute the Gilbert damping parameters as the magnetic moments is rotated from the *z* to *x* axis, as well as for magnetic moments rotating within the *x-o-y* plane.

## III. RESULTS AND DISCUSSIONS

Bulk FGT has the same hexagonal structure with a space group P6$_3$/mmc (No.194) as its sister vdW ferromagnetic material Fe$_3$GeTe$_2$. In each layer of bulk FGT, the Fe$_3$Ga plane is sandwiched by two Te layers along the *c* axis (i.e., the *z* axis), and two different types of Fe atoms occupy two different positions. In FGT bulk, its monolayers stack in an ABAB sequence along the *c* axis with weak vdW interactions. As shown in Fig. 1(a), the Ga atom in the upper layer is located directly above the Fe atom in the middle of the lower layer in bilayer FGT. We add 15 Å vacuum space to avoid effects between the adjacent bilayer FGT when constructing its slab model. For the investigation of the anisotropy of Gilbert damping, we define angles $\theta$ and $\varphi$ [Fig. 1(a)] between the *x* axis and magnetization when it is rotated in *z-o-x* plane and *x-o-y* plane, respectively. By calculating the MAE of bilayer FGT which is defined as $E_{MAE} = E_{in\text{-}plane} - E_{out\text{-}of\text{-}plane}$, we obtain that it is 0.24 meV/Fe. Hence, bilayer FGT also has a strong perpendicular magnetic anisotropy, similar to bulk FGT [15,66,67].

Fig. 1(c) shows the band structure of bilayer FGT along the high symmetry path of $\Gamma - K - M - \Gamma$ with its spin along its easy axis (i.e., the *z* axis) when SOC is included in our DFT calculations. We can observe that bilayer FGT exhibits metallic behavior and its projected band structure indicates that Fe atoms dominate the band contributions whereas Ga and Te atoms have minimal states near the Fermi level. As shown in Fig. 1(d), we also calculate the density of states (DOS) of bilayer FGT. The nonzero DOS at the Fermi level also confirms the metallic character of bilayer FGT. Throughout the energy range from -2.0 to 1.0 eV, the DOS of Fe atom is dominant while Te and Ga atoms contribute very little, matching our band structure. Combining the dominant contribution of Fe atoms at the Fermi level and their weak SOC, it is expected that a small Gilbert damping may occur in bilayer FGT when the magnetization is along the *z* axis.

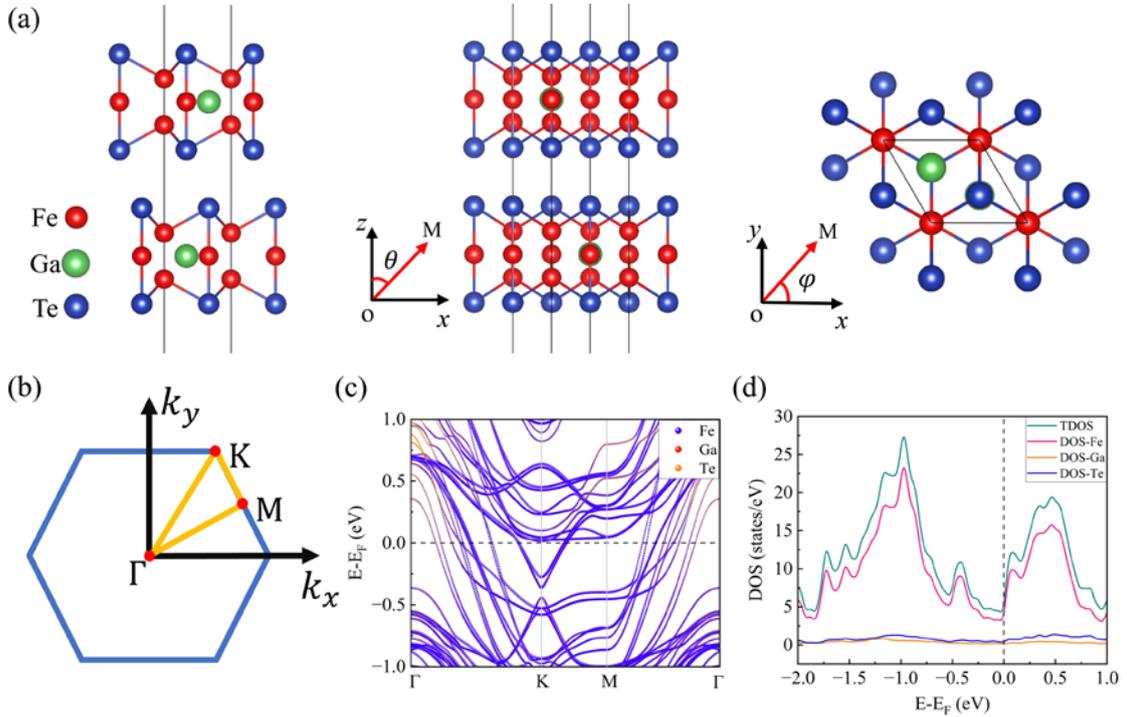

FIG. 1. (a) Side view (left and middle) and top view (right) of the crystal structure of bilayer FGT. The insets show the types of atom and define the magnetization orientation denoted as $\theta$ and $\varphi$. (b) The first Brillouin zone of bilayer FGT. DFT+SOC calculated (c) Atomically projected band structure and (d) projected DOS of bilayer FGT when its magnetization is along the *z* axis. In (c), the colors indicate the band weights of different atoms.

Considering the perpendicular magnetic anisotropy of bilayer FGT, we first study its Gilbert damping when its magnetization is along the *z* axis (i.e., $\theta = 0°$). As shown in Fig. 2(a), the Gilbert damping varies non-monotonically with respect to the scattering rate: it first decreases and then increases as the scattering rate rises, which is similar to the results reported in many previous studies [35,62,68]. The underlying physical mechanism of this non-monotonical trend can be elucidated by the breathing Fermi surface model [28]. The Gilbert damping in metallic ferromagnets is composed of two distinct contributions. One is the conductivity-like term, which increases as the relaxation time (the inverse of the scattering rate) increases. The other is the resistivity-like term, which decreases with increasing the relaxation time. Consequently, these two contributions with opposite trends with respect to the scattering rate lead to a non-monotonic Gilbert damping. It is worth noting that bilayer FGT has a fairly low Gilbert damping of $2.5×10^{-3}$ when the temperature is 300 K (i.e., the scattering rate $\Gamma = 26$ meV), which is consistent with our expectation acquired according to the projected band structure. Such low Gilbert damping suggests that bilayer FGT may be a potential playground for developing magnonics devices [69].

To explore the orientational anisotropy of the Gilbert damping in bilayer FGT, we calculate its Gilbert damping with its magnetization along different directions. To this end, we rotate its magnetization from the magnetic easy axis (i.e., the *z* axis) to the hard one (i.e., the *x* axis) within the *z-o-x* plane with an interval of five degrees and calculate the corresponding Gilbert damping. For the sake of clarity, the Gilbert dampings of a handful of salient angles, namely $\theta = 0°$, $30°$, $60°$ and $90°$ are shown in Fig. 2(a) and other results are given in Fig. S1(a) in the SM [59]. We see that the Gilbert dampings are different for each magnetization orientation, revealing that there is a noticeable orientational anisotropy in the Gilbert damping, especially in the low scattering rate zone. This phenomenon is attributed to the fact that the conductivity-like contribution displays a pronounced orientational anisotropy in the entire scattering rate range [28]. In contrast, the resistivity-like contribution exhibits a diminishing orientational anisotropy with increasing scattering rate due to the lifetime broadening [28].

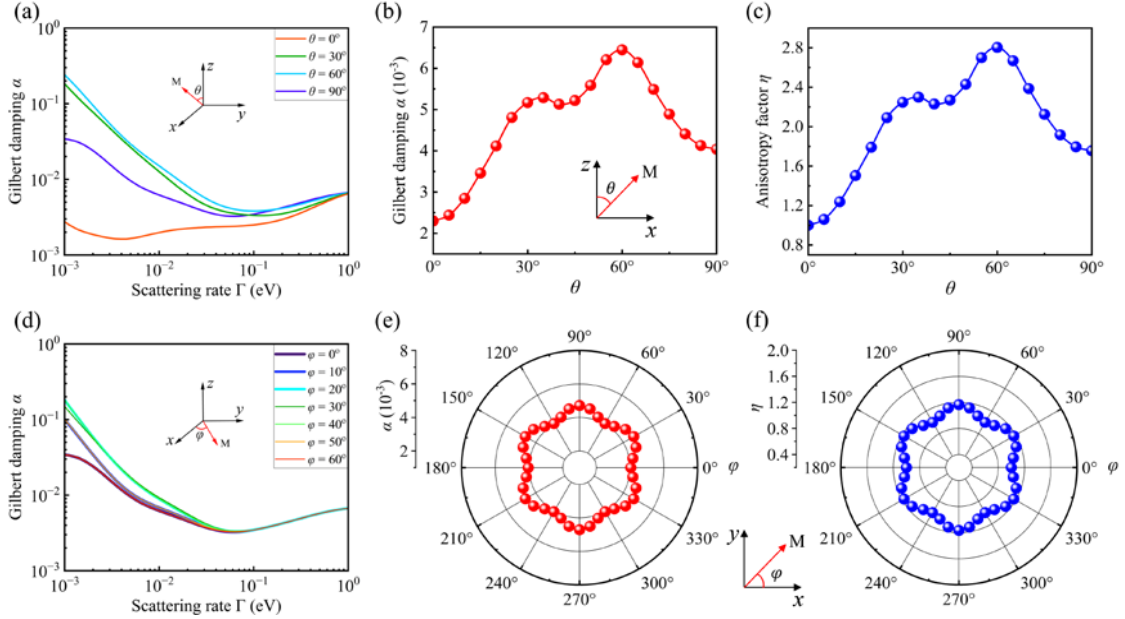

FIG. 2. Scattering rate dependent Gilbert damping of bilayer FGT with magnetization (a) varying from the $z$ axis ($\theta = 0°$) to $x$ axis ($\theta = 90°$) in intervals of 30°, (d) within the $x$-$o$-$y$ plane in intervals of 10°. $\theta$-dependent (b) Gilbert damping and (c) anisotropy factor of bilayer FGT at $\Gamma = 26$ meV (i.e., 300 K). $\varphi$-dependent (e) Gilbert damping and (f) anisotropy factor of bilayer FGT at $\Gamma = 26$ meV.

As spintronic devices operate at room temperature in practice and the $T_C$ of FGT exceeds 300 K, here we focus on the orientational anisotropy of the Gilbert damping at a scattering rate of 26 meV (corresponding to 300 K). Fig. 2(b) shows the Gilbert damping parameter $\alpha$ obtained at 300 K when the magnetization rotates from the $z$ to $x$ axis. Explicitly, $\alpha$ initially increases and subsequently decreases, and the maximum and minimum values of the Gilbert damping occur at $\theta = 60°$ and $\theta = 0°$, respectively. This indicates that the Gilbert damping of bilayer FGT exhibits a pronounced anisotropy within the $z$-$o$-$x$ plane. To intuitively quantify the anisotropy of the Gilbert damping, we introduce an anisotropy factor $\eta = \alpha/\alpha_{min}$ where $\alpha_{min}$ represents the minimum value of the Gilbert damping. Note that such definition of $\eta$ is inspired by the previous study on the Gilbert damping anisotropy of CoFe films [33]. As shown in Fig. 2(c), $\eta$ reaches its peak at $\theta = 60°$, with a remarkable magnitude of approximately 2.8. In a previous study [70], the anisotropic Gilbert damping of $Fe_3GeTe_2$ monolayer was also explored. When the magnetization direction sweeps from the $x$ axis to the $z$ axis and then to the -$x$ axis, the Gilbert damping of $Fe_3GeTe_2$ monolayer follows a trend of the cosine function and reaches its maximum and minimum values at $x$ axis (-$x$ axis) and $z$ axis, respectively. This is similar to the minimum Gilbert damping of bilayer FGT at $\theta = 0°$, whereas the anisotropy factor of Gilbert damping in $Fe_3GeTe_2$ monolayer is 1.62 and smaller than that of bilayer FGT.

Here, we investigate the Gilbert damping of bilayer FGT with its magnetization altering in the *x-o-y* plane. Given the in-plane six-fold improper rotation symmetry of bilayer FGT, we illustrate its Gilbert dampings in Fig. 2(d) when its magnetization is oriented with the in-plane angle $\varphi$ from 0° to 60°. Similar to the case of magnetization along the magnetic easy axis, the Gilbert dampings for the case of magnetizations in the *x-o-y* plane are non-monotonical as the scattering rate increases. Although we observe the $\Gamma$-dependent Gilbert dampings for different magnetic moment directions, only four distinct curves of Gilbert dampings are evident. This suggests that Gilbert dampings exhibit almost the same behavior at some certain angles. We extend our calculations of Gilbert dampings for the in-plane angle $\varphi$ covering the full 0°-360° range. As shown Figs. S1(b)-S1(f) in the SM [59], we still only see four curves for Gilbert damping and the Gilbert dampings display the expected six-fold symmetry regarding the in-plane angle $\varphi$. Then, we extract the Gilbert damping parameter $\alpha$ for each in-plane angle at room temperature and calculate the anisotropy factor $\eta$. Due to the intrinsic crystal structure of bilayer FGT, the Gilbert dampings present a six-fold symmetry within the *x-o-y* plane in Fig. 2(e). Simultaneously, Fig. 2(f) depicts the maximum anisotropy factor $\eta_{max} = 1.14$, which means the largest variation of Gilbert dampings is only 14% (at $\varphi = 30°$). So, the Gilbert dampings of bilayer FGT are weakly anisotropic when its magnetization is in-plane.

As given by the Eq. (2), the expression for Gilbert damping incorporates two delta functions depending on the energy differences with respect to the Fermi level (i.e., $(E_i - E_F)$ and $(E_j - E_F)$). This indicates that only those bands that traverse the Fermi surface or possess energies close to the Fermi level can make an important contribution to the Gilbert damping. If the bands are distant from the Fermi level, their contributions become imperceptible. In addition, the Hamiltonian in Eq. (2) also encompasses terms related with SOC. When the direction of the magnetization changes, the SOC usually alters accordingly, which in turn affects the magnitude of the Gilbert damping. To unravel the intricate physical mechanism of anisotropic Gilbert dampings, we utilize DFT+SOC calculations to investigate the band structures at high-symmetry paths under some specific magnetization orientations, such as $\varphi = 0°$, 30° and $\theta = 0°$, 60°, 90°. As shown in Figs. 3(a)-3(b), when the magnetization lies within the *x-o-y* plane, the bands at $\varphi = 0°$ and $\varphi = 30°$ appear basically indistinguishable. Yet, upon closer inspection on the bands near the Fermi level, subtle differences begin to emerge. It is these minute distinctions in the bands that give rise to the very weak anisotropic Gilbert dampings. When the magnetization is rotated from the *z* to *x* axis, the bands at $\theta = 0°$, 60°, 90° show obvious differences originating from the impact of magnetization orientations and the zoomed views also reveal marked different bands which intersect the Fermi level as illustrated in Fig. 3(b). Hence, these salient differences in the bands lead to the strong anisotropic Gilbert dampings when the magnetization of bilayer FGT is rotated from the z axis to in-plane.

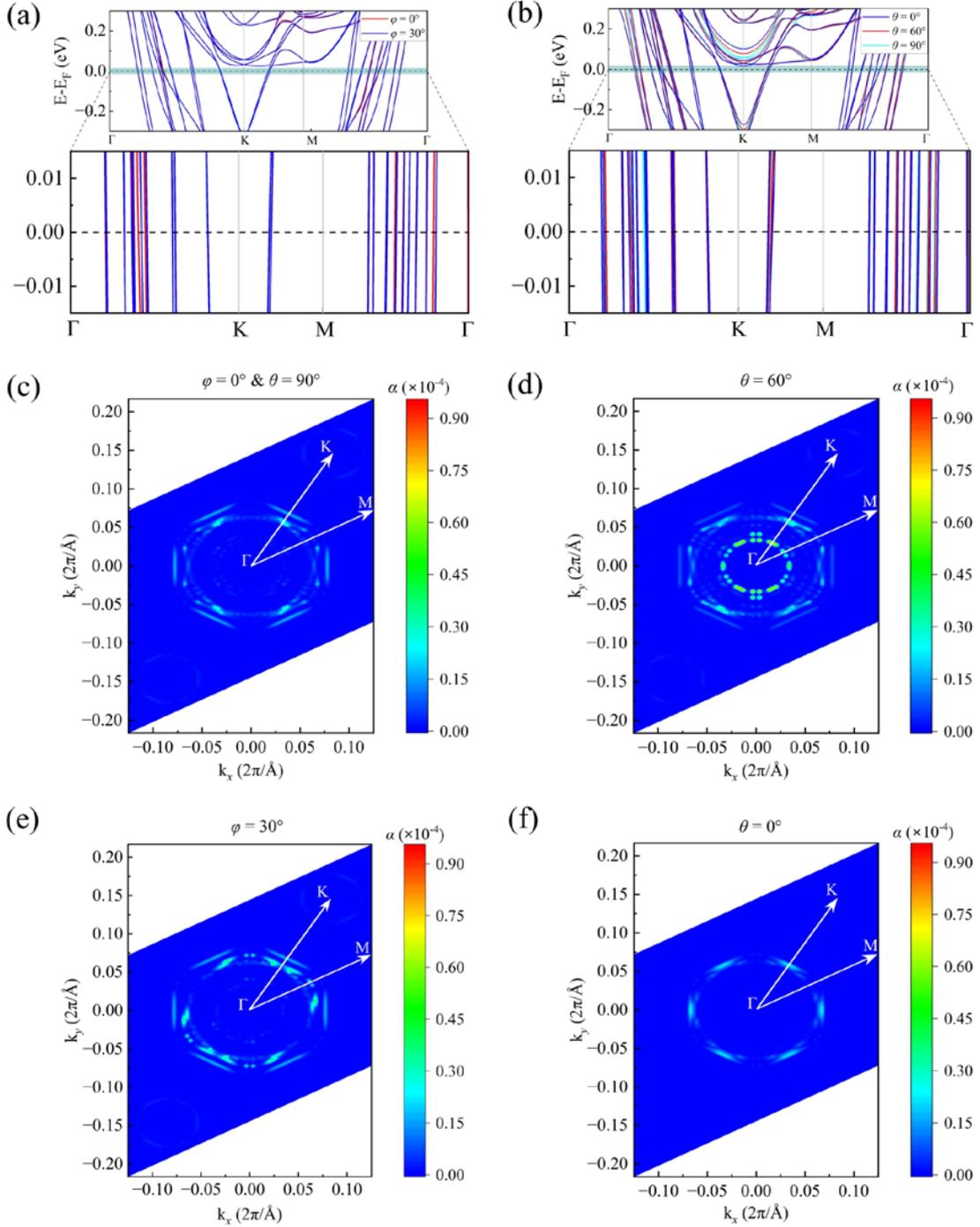

FIG. 3. DFT+SOC calculated band structures of bilayer FGT at (a) $\varphi = 0°$ (red lines) and $\varphi = 30°$ (blue lines), (b) $\theta = 0°$ (blue lines), $\theta = 60°$ (red lines) and $\theta = 90°$ (cyan lines). The $k$-dependent Gilbert damping contributions in the first Brillouin zone to the Gilbert damping of bilayer FGT for (c) $\varphi = 0°$ (& $\theta = 90°$), (d) $\theta = 60°$, (e) $\varphi = 30°$, (f) $\theta = 0°$. The color bars on the right of (c)-(f) indicate the values of Gilbert damping contributions from the highest to the lowest.

To obtain an insight into the effect of magnetization orientations on the Gilbert dampings, we examine the $k$-dependent Gilbert damping contributions along the $\Gamma-$ $K-M-\Gamma$ path for some specific magnetization orientations (Fig. S2 in the SM [59]). By comparing the $k$-dependent contributions with the bands, we discern that the $k$-points whose bands intersect or locate near the Fermi level play a pivotal role in contributing to the Gilbert dampings, with the $\Gamma-K$ path being particularly significant. It is worth pointing out that when the magnetization is rotated from the $z$ to $x$ axis, the maximal contribution of $\theta=0°$ is substantially lower than that of $\theta=60°$. This finding is consistent with our previous results. For the two in-plane orientations, namely $\varphi=0°$, and $\varphi=30°$, their peaks manifest at very similar positions. Nevertheless, the maximal peak at $\varphi=0°$ surpasses that at $\varphi=30°$, which contradicts the occurrence of the in-plane maximum Gilbert damping at $\varphi=30°$. We infer that this discrepancy could be attributed to the non-negligible contributions of $k$-points beyond the high-symmetry paths to the Gilbert damping, thereby accounting for the observed differences.

In addition, we perform a comprehensive investigation of the $k$-dependent Gilbert damping contributions in the first Brillouin zone and illustrate the results in Figs. 3(c)-3(f). Overall, there exist stark disparities among different magnetization orientations. At $\theta=0°$, there are only six $k$-point zones where the Gilbert damping contributions are moderate. As the magnetization is rotated to the in-plane, we see that $k$-point zones even with the highest Gilbert damping contribution appear at $\theta=60°$. Compared to the case of $\theta=0°$, the Gilbert damping contribution in the case of $\theta=90°$ displays a handful of additional peaks. Furthermore, we can clearly observe that the pattern of the Gilbert damping contribution at $\varphi=30°$ is more complex and larger than that at $\varphi=0°$, indicating that extra $k$-points make non-negligible contribution to the Gilbert damping. This explains why the Gilbert damping is elevated at $\varphi=30°$ in comparison to that at $\varphi=0°$. Besides, we mark the high-symmetry points in Figs. 3(c)-3(f). There are indeed peaks along the $\Gamma-K$ path, which is more noticeable than along the $\Gamma-M$ path, which matches well with the results above mentioned. To make the results clearer, we also provide three-dimensional illustrations of the $k$-point dependent Gilbert damping contributions in the first Brillouin zone (Fig. S3 in the SM [59]).

Owing to the extra twist degree of freedom that presents in a bilayer structure and considering the computational resources required, we execute a twist of 43.9° to bilayer FGT. Such twist yields a twisted bilayer FGT which contains 150 atoms in its unit cell, as depicted in Fig. 4(a). Similarly, we denote the angles between the $x$ axis (i.e., the $a$ axis) and the magnetization orientation when it points in $z$-$o$-$x$ plane and $x$-$o$-$y$ plane as $\theta$ and $\varphi$, respectively. The MAE (i.e., $E_{in\text{-}plane}-E_{out\text{-}of\text{-}plane}$) of the twisted bilayer FGT is 0.26 meV/Fe, which is still perpendicular and slightly larger than that of bilayer FGT. To obtain a preliminary understanding on the effect of the twist on the Gilbert damping of the bilayer FGT, we first calculate the Gilbert dampings of the twisted bilayer FGT when its magnetization is aligned with the *x* and *z* axis, respectively, and compare the results with those of bilayer FGT. As shown in Fig. 4(b), the overall trend of the Gilbert damping of the twisted bilayer FGT is also non-monotonic as the scattering rate grows, similar to bilayer FGT. However, it is worth mentioning that at room temperature (corresponding to $\Gamma = 26$ meV), the in-plane Gilbert damping of the twisted bilayer FGT increases by 110% while the out-of-plane one decreases by 36%, compared to those of bilayer FGT.

We now study the effect of twist on the anisotropy of the Gilbert dampings of the twisted bilayer FGT. As shown in Fig. 4(c), there appear only two visible curves with very close trends when the magnetization rotates within the *x-o-y* plane. This indicates that when the magnetization is in-plane, the Gilbert damping is almost isotropic over the entire scattering rate range. We extract the Gilbert damping of the twisted bilayer FGT under each magnetization orientation at 300 K and compute the corresponding anisotropy factor. As shown in Fig. 4(d), the Gilbert damping within the *x-o-y* plane in the twisted bilayer FGT is notably enhanced compared to bilayer FGT. The introduction of twist leads to a reduction in interfacial symmetry, which in turn strengthens the SOC [71]. Because the Gilbert damping is directly related to SOC, such strengthening results in an enhanced in-plane Gilbert damping. Besides, we observe the nearly constant $\alpha$ across all directions in Fig. 4(d) and the maximum variation of Gilbert damping is merely 0.2% in Fig. 4(e). This phenomenon may arise from the twist-induced average effects of various in-plane directions, which makes the Gilbert damping more isotropic.

Figs. 4(f)-4(h) show the Gilbert dampings of the twisted bilayer FGT when its magnetization is rotated from the *z* to *x* axis. As can be seen in Fig. 4(f), there exists an obvious orientational anisotropy in the Gilbert dampings throughout the nearly entire scattering rate range. This is fully different from the almost vanishing orientational anisotropy of the Gilbert damping for the in-plane magnetizations. In order to facilitate a more intuitive comparison with bilayer FGT, we examine the Gilbert dampings and the anisotropy factor $\eta$ for each magnetization orientation at 300 K. Similar to bilayer FGT, the twisted bilayer FGT also witnesses its maximum and minimum Gilbert damping at $\theta = 60°$ and $\theta = 0°$. Astonishingly, the twisted bilayer FGT shows a much stronger anisotropy. During the variation of magnetization orientations in the *z-o-x* plane, the anisotropy factor $\eta$ soars to an impressive peak of 6.35, which is the largest reported so far and more than double that of bilayer FGT. This indicates that a twist of 43.9° can significantly enhance the orientational anisotropy of Gilbert damping within *z-o-x* plane in bilayer FGT, highlighting its essential role in adjusting this critical magnetic property.

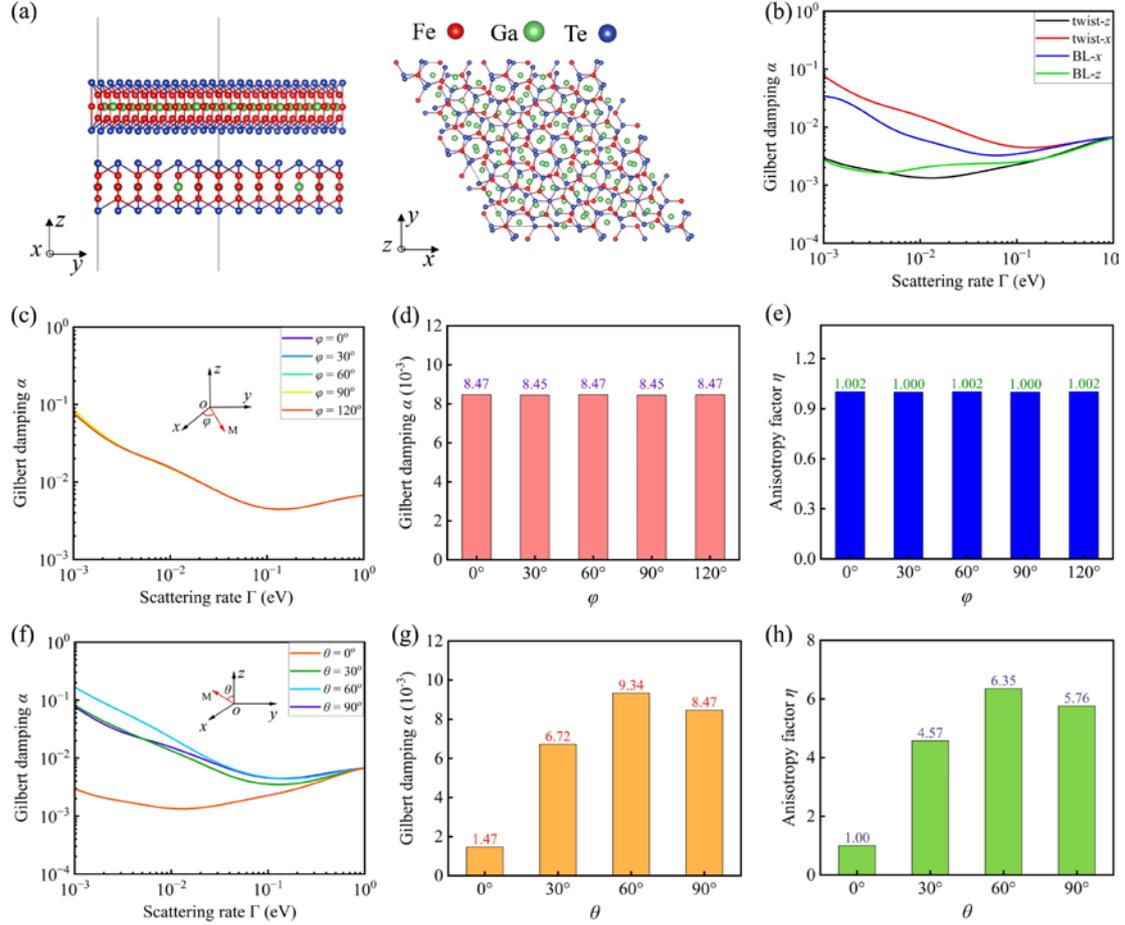

FIG. 4. (a) Side (left) and top (right) view of the crystal structure of the twisted bilayer FGT. (b) Comparison of the Gilbert dampings between bilayer and twisted bilayer FGT with their magnetization along the *x* and *z* axis. Scattering rate ($\Gamma$) dependent Gilbert damping of the twisted bilayer FGT with its magnetization changing (c) in *x-o-y* plane at intervals of 30°, and (f) from *z* to *x* axis in intervals of 30°. $\varphi$-dependent (d) Gilbert damping and (e) anisotropy factor of the twisted bilayer FGT at $\Gamma = 26$ meV (300 K). $\theta$-dependent (g) Gilbert damping and (h) anisotropy factor of the twisted bilayer FGT at $\Gamma = 26$ meV (300 K).

To illuminate the underlying mechanism for the highly tunable Gilbert damping anisotropy as observed in the twisted bilayer FGT, we study its band structures with SOC at three typical angles $\varphi = 0°$ & $\theta = 90°$, $\theta = 60°$, and $\theta = 0°$, as shown in Figs. 5(a)-5(c). Given that the twisted bilayer FGT has 150 atoms and a reduced symmetry, its band structure is considerably complex compared with bilayer FGT. Meanwhile, the differences between the bands structures of the twisted bilayer FGT with varied magnetization orientations are more obvious than those between the band structures of bilayer FGT with different magnetization orientations. As depicted in Figs. S4(a)-S4(c) in the SM [59], the zoomed views near the Fermi level highlight these differences. Such magnetization orientation dependent band structures may potentially foreshow the strongly anisotropic Gilbert dampings of the twisted bilayer FGT. We carry out an in-depth analysis of the *k*-point contributions to the Gilbert dampings along the high-symmetry path $\Gamma - K - M - \Gamma$. As shown in Figs. 5(d)-5(f), upon aligning the magnetization along the *x* axis, conspicuous peaks appear around the $\Gamma$ point and along the $\Gamma - K$ path, coinciding with some bands in proximity to or intersecting the Fermi level. The most prominent peak emerges on the $\Gamma - K$ path, while several smaller extrema also appear near the $\Gamma$, K and M points at $\theta = 60°$. When the magnetization is oriented along the *z* axis, the Gilbert damping contributions is much smaller than those in the case of $\theta = 90°$ and $\theta = 60°$.

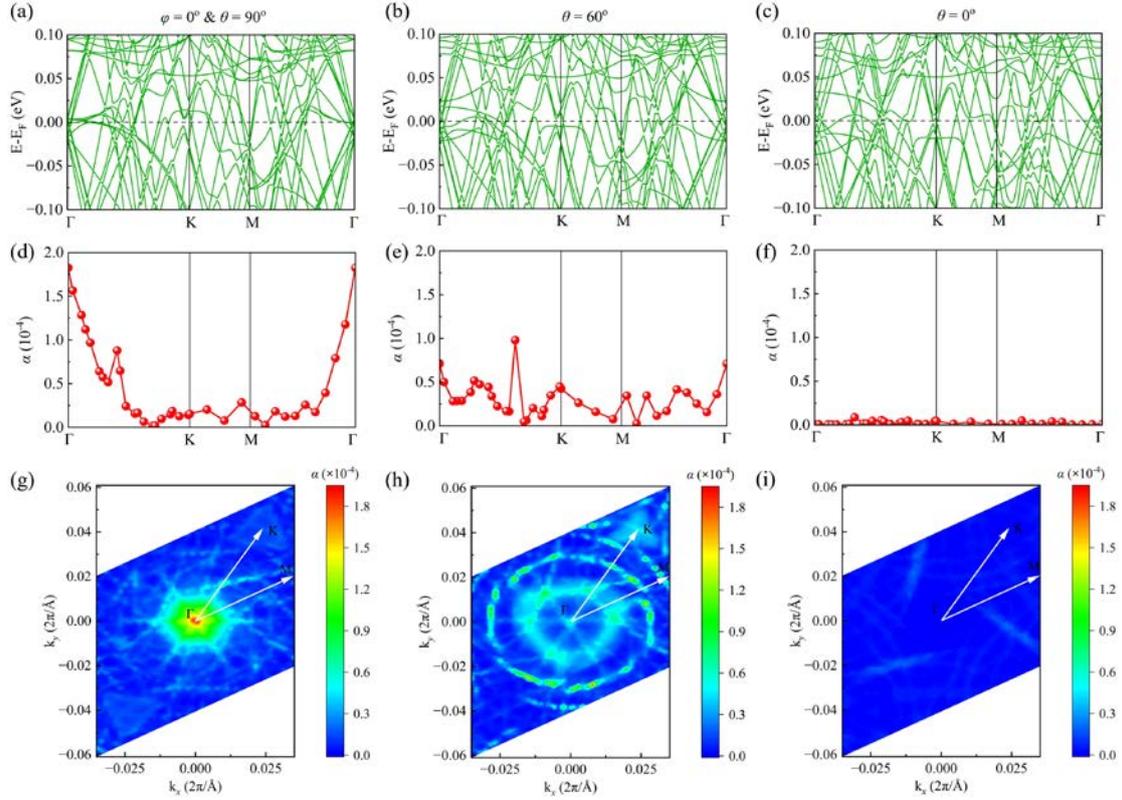

FIG. 5. DFT+SOC calculated band structures and *k*-dependent Gilbert dampings of the twisted bilayer FGT for (a, d) $\varphi = 0°$ (& $\theta = 90°$), (b, e) $\theta = 60°$, (c, f) $\theta = 0°$. The *k*-dependent Gilbert damping contributions in the first Brillouin zone of the twisted bilayer FGT for (g) $\varphi = 0°$ (& $\theta = 90°$), (h) $\theta = 60°$, (i) $\theta = 0°$. The color bars on the right of (g)-(i) indicate the values of Gilbert damping contributions from the highest to the lowest.

If we solely consider the contributions of *k*-points along the high-symmetry paths, $\varphi = 0°$ manifests larger peaks than $\theta = 60°$. For a holistic picture, we compute the *k*-point contributions to the Gilbert damping of twisted bilayer FGT throughout the entire first Brillouin zone and show the results in Figs. 5(g)-5(i). For $\varphi = 0°$, the nearly

Page 13 of 19

hexagonal *k*-point zone around the center of the first Brillouin zone makes dominant contributions to the Gilbert damping. By marking the high-symmetry points Γ, K and M, one sees that the Gilbert damping contribution reaches its maximum at the Γ point and gradually decreases along Γ − K and Γ − M paths. When the magnetization is rotated from the *z* to *x* axis by an angle of $\theta = 60°$, a significant portion of *k* points in the Brillouin zone actively contribute to the Gilbert damping. This is the reason why the Gilbert damping at $\theta = 60°$ is slightly enhanced compared with that at $\varphi = 0°$. Consistent with the results shown in Fig. 5(e), sizable Gilbert damping contributions are present near the Γ, K and M points as shown in Fig. 5(h). As the magnetization is aligned perpendicular to the *x-o-y* plane (i.e., $\theta = 0°$), the Gilbert damping contributions are very small, and only some *k*-point zones exhibit minor contributions. This illustrates the very low Gilbert damping of $1.47×10^{-3}$ at $\theta = 0°$. The distinct contrasts between the *k*-dependent Gilbert damping contributions maps (also see Fig. S5 in the SM [59] for the three-dimensional illustrations) at these three typical angles reaffirm our observation of the pronounced Gilbert damping anisotropy in the twisted bilayer FGT when the magnetization orientation evolves from the *z* axis to in-plane.

## IV. CONCLUSIONS

To sum up, we systematically study the Gilbert damping of 2D vdW ferromagnets, bilayer FGT and the twisted bilayer FGT, using first-principles calculations. When the magnetization is along the *z* axis, we obtain a fairly low Gilbert damping in bilayer FGT and twist can further reduce it in the twisted bilayer FGT. When the magnetization is in the *x-o-y* plane, the twisted bilayer FGT has an enhanced but much more isotropic Gilbert damping than bilayer FGT. Such different Gilbert dampings are related with the strengthened SOC resulting from the lower interfacial symmetry and the average effect in the twisted bilayer FGT. We find a large orientational anisotropy of Gilbert damping in bilayer FGT when its magnetization is rotated from the *z* to *x* axis. Surprisingly, the anisotropy factor can be enhanced to be unprecedently large (i.e., 6.35) in the twisted bilayer FGT. We illustrate that the magnetization orientation dependent SOC leads to anisotropic band structures, giving rise to Gilbert damping anisotropy in both bilayer and the twisted bilayer FGT. Given the twisting-induced pronounced effect on Gilbert damping, bilayer FGT with smaller twist angles which may have qualitatively different magnetic behavior deserves future explorations by means of combining DFT and the rapidly developed machine learning techniques. Our study provides an in-depth study of the impact of twist on the Gilbert damping in 2D vdW ferromagnets, offering a new concept for modulating the Gilbert damping in the design of spintronic devices.

**Acknowledgements**

This work was supported by the National Key R&D Program of China (Grant No. 2024YFA1408303 and 2022YFA1403301), the National Natural Sciences Foundation of China (Grants No. 12474247, 92165204). Yusheng Hou acknowledges the support from Guangdong Provincial Key Laboratory of Magnetoelectric Physics and Devices (Grant No. 2022B1212010008) and Research Center for Magnetoelectric Physics of Guangdong Province (Grant No. 2024B0303390001). Density functional theory calculations are performed on Tianhe-Xingyi.

**References**

[1] K. S. Novoselov, A. K. Geim, S. V. Morozov, D. Jiang, Y. Zhang, S. V. Dubonos, I. V. Grigorieva, and A. A. Firsov, Electric field effect in atomically thin carbon films, Science **306**, 666 (2004).

[2] K. S. Novoselov, D. Jiang, F. Schedin, T. Booth, V. Khotkevich, S. Morozov, and A. K. Geim, Two-dimensional atomic crystals, PNAS **102**, 10451 (2005).

[3] K. S. Novoselov, A. Mishchenko, A. Carvalho, and A. Castro Neto, 2D materials and van der Waals heterostructures, Science **353**, aac9439 (2016).

[4] L. Xie, Two-dimensional transition metal dichalcogenide alloys: preparation, characterization and applications, Nanoscale **7**, 18392 (2015).

[5] A. K. Geim and I. V. Grigorieva, Van der Waals heterostructures, Nature **499**, 419 (2013).

[6] K. F. Mak, C. Lee, J. Hone, J. Shan, and T. F. Heinz, Atomically thin $MoS_2$: a new direct-gap semiconductor, Phys. Rev. Lett. **105**, 136805 (2010).

[7] N. D. Mermin and H. Wagner, ABSENCE OF FERROMAGNETISM OR ANTIFERROMAGNETISM IN ONE- OR TWO-DIMENSIONAL ISOTROPIC HEISENBERG MODELS, Phys. Rev. Lett. **17**, 1133 (1966).

[8] C. Gong *et al.*, Discovery of intrinsic ferromagnetism in two-dimensional van der Waals crystals, Nature **546**, 265 (2017).

[9] B. Huang *et al.*, Layer-dependent ferromagnetism in a van der Waals crystal down to the monolayer limit, Nature **546**, 270 (2017).

[10] Z. Fei *et al.*, Two-dimensional itinerant ferromagnetism in atomically thin $Fe_3GeTe_2$, Nat. Mater. **17**, 778 (2018).

[11] Y. Deng *et al.*, Gate-tunable room-temperature ferromagnetism in two-dimensional $Fe_3GeTe_2$, Nature **563**, 94 (2018).

[12] W. Chen, Z. Sun, Z. Wang, L. Gu, X. Xu, S. Wu, and C. Gao, Direct observation of van der Waals stacking–dependent interlayer magnetism, Science **366**, 983 (2019).

[13] D. J. O'Hara *et al.*, Room temperature intrinsic ferromagnetism in epitaxial manganese selenide films in the monolayer limit, Nano Lett. **18**, 3125 (2018).

[14] M. M. Otrokov *et al.*, Prediction and observation of an antiferromagnetic topological insulator, Nature **576**, 416 (2019).

[15] G. Zhang, F. Guo, H. Wu, X. Wen, L. Yang, W. Jin, W. Zhang, and H. Chang, Above-room-temperature strong intrinsic ferromagnetism in 2D van der Waals $Fe_3GaTe_2$ with large perpendicular magnetic anisotropy, Nat. Commun. **13**, 5067 (2022).

[16] T. L. Gilbert, A phenomenological theory of damping in ferromagnetic materials, IEEE Trans. Magn. **40**, 3443 (2004).

[17] V. L. Safonov, Tensor form of magnetization damping, J. Appl. Phys. **91**, 8653 (2002).

[18] D. Steiauf and M. Fähnle, Damping of spin dynamics in nanostructures: An ab initio study, Phys. Rev. B **72**, 064450 (2005).

[19] M. Fähnle and D. Steiauf, Breathing Fermi surface model for noncollinear magnetization: A generalization of the Gilbert equation, Phys. Rev. B **73**, 184427 (2006).

[20] S. Azzawi, A. T. Hindmarch, and D. Atkinson, Magnetic damping phenomena in

ferromagnetic thin-films and multilayers, J. Phys. D: Appl. Phys. **50**, 473001 (2017).

[21] O. Ertl, G. Hrkac, D. Suess, M. Kirschner, F. Dorfbauer, J. Fidler, and T. Schrefl, Multiscale micromagnetic simulation of giant magnetoresistance read heads, J. Appl. Phys. **99** (2006).

[22] N. Smith, Micromagnetic modeling of magnoise in magnetoresistive read sensors, J. Magn. Magn. Mater. **321**, 531 (2009).

[23] S. S. P. Parkin, M. Hayashi, and L. Thomas, Magnetic Domain-Wall Racetrack Memory, Science **320**, 190 (2008).

[24] A. Brataas, A. D. Kent, and H. Ohno, Current-induced torques in magnetic materials, Nat. Mater. **11**, 372 (2012).

[25] D. C. Ralph and M. D. Stiles, Spin transfer torques, J. Magn. Magn. Mater. **320**, 1190 (2008).

[26] K. Lee and S. H. Kang, Development of Embedded STT-MRAM for Mobile System-on-Chips, IEEE Trans. Magn. **47**, 131 (2011).

[27] M. Fähnle, D. Steiauf, and J. Seib, The Gilbert equation revisited: anisotropic and nonlocal damping of magnetization dynamics, J. Phys. D: Appl. Phys. **41**, 164014 (2008).

[28] K. Gilmore, M. Stiles, J. Seib, D. Steiauf, and M. Fähnle, Anisotropic damping of the magnetization dynamics in Ni, Co, and Fe, Phys. Rev. B **81**, 174414 (2010).

[29] J. Seib, D. Steiauf, and M. Fähnle, Linewidth of ferromagnetic resonance for systems with anisotropic damping, Phys. Rev. B **79**, 092418 (2009).

[30] D. Thonig and J. Henk, Gilbert damping tensor within the breathing Fermi surface model: anisotropy and non-locality, New J. Phys. **16**, 013032 (2014).

[31] X. Yang *et al.*, Anisotropic Nonlocal Damping in Ferromagnet/α-GeTe Bilayers Enabled by Splitting Energy Bands, Phys. Rev. Lett. **131**, 186703 (2023).

[32] L. Chen *et al.*, Emergence of anisotropic Gilbert damping in ultrathin Fe layers on GaAs (001), Nat. Phys. **14**, 490 (2018).

[33] Y. Li *et al.*, Giant anisotropy of Gilbert damping in epitaxial CoFe films, Phys. Rev. Lett. **122**, 117203 (2019).

[34] H. Xu, H. Chen, F. Zeng, J. Xu, X. Shen, and Y. Wu, Giant anisotropic Gilbert damping in single-crystal Co-Fe-B (001) Films, Phys. Rev. Appl. **19**, 024030 (2023).

[35] S.-B. Zhao, X.-F. Huang, Z.-q. Wang, R. Wu, and Y. Hou, Thickness-dependent anisotropic Gilbert damping in heterostructures of ferromagnets and two-dimensional ferroelectric bismuth monolayers, Phys. Rev. B **110**, 094435 (2024).

[36] S. K. Behura, A. Miranda, S. Nayak, K. Johnson, P. Das, and N. R. Pradhan, Moiré physics in twisted van der Waals heterostructures of 2D materials, Emergent Materials **4**, 813 (2021).

[37] J. M. B. Lopes dos Santos, N. M. R. Peres, and A. H. Castro Neto, Graphene Bilayer with a Twist: Electronic Structure, Phys. Rev. Lett. **99**, 256802 (2007).

[38] E. Suárez Morell, J. D. Correa, P. Vargas, M. Pacheco, and Z. Barticevic, Flat bands in slightly twisted bilayer graphene: Tight-binding calculations, Phys. Rev. B **82**, 121407 (2010).

[39] S. Lisi *et al.*, Observation of flat bands in twisted bilayer graphene, Nat. Phys. **17**, 189 (2021).

[40] R. Bistritzer and A. H. MacDonald, Moiré bands in twisted double-layer graphene, PNAS **108**, 12233 (2011).

[41] A. J. Jones *et al.*, Observation of Electrically Tunable van Hove Singularities in Twisted Bilayer Graphene from NanoARPES, Adv. Mater. **32**, 2001656 (2020).

[42] G. Li, A. Luican, J. M. B. Lopes dos Santos, A. H. Castro Neto, A. Reina, J. Kong, and E. Y. Andrei, Observation of Van Hove singularities in twisted graphene layers, Nat. Phys. **6**, 109 (2010).

[43] Y. Cao, V. Fatemi, S. Fang, K. Watanabe, T. Taniguchi, E. Kaxiras, and P. Jarillo-Herrero, Unconventional superconductivity in magic-angle graphene superlattices, Nature **556**, 43 (2018).

[44] Y. Cao *et al.*, Correlated insulator behaviour at half-filling in magic-angle graphene superlattices, Nature **556**, 80 (2018).

[45] X. Lu *et al.*, Superconductors, orbital magnets and correlated states in magic-angle bilayer graphene, Nature **574**, 653 (2019).

[46] A. L. Sharpe, E. J. Fox, A. W. Barnard, J. Finney, K. Watanabe, T. Taniguchi, M. Kastner, and D. Goldhaber-Gordon, Emergent ferromagnetism near three-quarters filling in twisted bilayer graphene, Science **365**, 605 (2019).

[47] M. Serlin, C. L. Tschirhart, H. Polshyn, Y. Zhang, J. Zhu, K. Watanabe, T. Taniguchi, L. Balents, and A. F. Young, Intrinsic quantized anomalous Hall effect in a moiré heterostructure, Science **367**, 900 (2020).

[48] K. P. Nuckolls, M. Oh, D. Wong, B. Lian, K. Watanabe, T. Taniguchi, B. A. Bernevig, and A. Yazdani, Strongly correlated Chern insulators in magic-angle twisted bilayer graphene, Nature **588**, 610 (2020).

[49] Y. Xie *et al.*, Fractional Chern insulators in magic-angle twisted bilayer graphene, Nature **600**, 439 (2021).

[50] R. Obata *et al.*, Pseudotunnel Magnetoresistance in Twisted van der Waals $Fe_3GeTe_2$ Homojunctions, Adv. Mater., 2411459 (2025).

[51] G. Kresse and J. Hafner, Ab initio molecular dynamics for liquid metals, Phys. Rev. B **47**, 558 (1993).

[52] G. Kresse and J. Furthmüller, Efficiency of ab-initio total energy calculations for metals and semiconductors using a plane-wave basis set, Comput. Mater. Sci **6**, 15 (1996).

[53] G. Kresse and J. Furthmüller, Efficient iterative schemes for ab initio total-energy calculations using a plane-wave basis set, Phys. Rev. B **54**, 11169 (1996).

[54] G. Kresse and D. Joubert, From ultrasoft pseudopotentials to the projector augmented-wave method, Phys. Rev. B **59**, 1758 (1999).

[55] J. P. Perdew, K. Burke, and M. Ernzerhof, Generalized gradient approximation made simple, Phys. Rev. Lett. **77**, 3865 (1996).

[56] J. P. Perdew and A. Zunger, Self-interaction correction to density-functional approximations for many-electron systems, Phys. Rev. B **23**, 5048 (1981).

[57] S. Grimme, Semiempirical GGA-type density functional constructed with a long-range dispersion correction, J. Comput. Chem. **27**, 1787 (2006).

[58] S. Grimme, J. Antony, S. Ehrlich, and H. Krieg, A consistent and accurate ab initio parametrization of density functional dispersion correction (DFT-D) for the 94 elements

H-Pu, The Journal of chemical physics **132** (2010).

[59] See Supplemental Material for (1) the comparison of structural and magnetic properties of bulk FGT under PBE (combing with DFT-D2 or DFT-D3) and LDA. (2) other twist angles and relevant data. (3) details of Gilbert damping calculations. (4) scattering rate dependent Gilbert dampings of bilayer FGT with magnetization orientation rotates within z-o-x and x-o-y plane. (5) band structures and k-dependent contributions to Gilbert dampings of bilayer FGT at representative angles. (6) 3D views of the k-dependent contributions of Gilbert damping of bilayer FGT in the 1BZ. (7) zoomed views of band structures near the Fermi level of twisted bilayer FGT at representative angles. (8) 3D views of the k-dependent contributions of Gilbert damping of twisted bilayer FGT in the 1BZ.

[60] X. Wang, R. Wu, D.-s. Wang, and A. Freeman, Torque method for the theoretical determination of magnetocrystalline anisotropy, Phys. Rev. B **54**, 61 (1996).

[61] J. Hu and R. Wu, Control of the Magnetism and Magnetic Anisotropy of a Single-Molecule Magnet with an Electric Field, Phys. Rev. Lett. **110**, 097202 (2013).

[62] Y. Hou and R. Wu, Strongly enhanced Gilbert damping in 3 d transition-metal ferromagnet monolayers in contact with the topological insulator $Bi_2Se_3$, Phys. Rev. Appl. **11**, 054032 (2019).

[63] A. Brataas, Y. Tserkovnyak, and G. E. Bauer, Scattering theory of Gilbert damping, Phys. Rev. Lett. **101**, 037207 (2008).

[64] A. Brataas, Y. Tserkovnyak, and G. E. Bauer, Magnetization dissipation in ferromagnets from scattering theory, Phys. Rev. B **84**, 054416 (2011).

[65] S. Mankovsky, D. Ködderitzsch, G. Woltersdorf, and H. Ebert, First-principles calculation of the Gilbert damping parameter via the linear response formalism with application to magnetic transition metals and alloys, Phys. Rev. B **87**, 014430 (2013).

[66] X. Li, M. Zhu, Y. Wang, F. Zheng, J. Dong, Y. Zhou, L. You, and J. Zhang, Tremendous tunneling magnetoresistance effects based on van der Waals room-temperature ferromagnet Fe3GaTe2 with highly spin-polarized Fermi surfaces, Appl. Phys. Lett. **122** (2023).

[67] A. M. Ruiz, D. L. Esteras, D. López-Alcalá, and J. J. Baldoví, On the Origin of the Above-Room-Temperature Magnetism in the 2D van der Waals Ferromagnet Fe3GaTe2, Nano Lett. **24**, 7886 (2024).

[68] K. Gilmore, Y. Idzerda, and M. D. Stiles, Identification of the Dominant Precession-Damping Mechanism in Fe, Co, and Ni by First-Principles Calculations, Phys. Rev. Lett. **99**, 027204 (2007).

[69] P. Pirro, V. I. Vasyuchka, A. A. Serga, and B. Hillebrands, Advances in coherent magnonics, Nature Reviews Materials **6**, 1114 (2021).

[70] P. Yang, R. Liu, Z. Yuan, and Y. Liu, Magnetic damping anisotropy in the two-dimensional van der Waals material $Fe_3GeTe_2$ from first principles, Phys. Rev. B **106**, 134409 (2022).

[71] C. Zhou, F. Kandaz, Y. Cai, C. Qin, M. Jia, Z. Yuan, Y. Wu, and Y. Ji, Anisotropic spin relaxation induced by surface spin-orbit effects, Phys. Rev. B **96**, 094413 (2017).